\documentstyle[epsbox]{ioplppt}
\def\D{{\cal D}}
\def\T{{\cal T}}

\begin{document}

\jl{1}

\title{$N$ Soliton Solutions to The Bogoyavlenskii-Schiff 
Equation and A Quest for The Soliton Solution in ($3 + 1$) 
Dimensions}[Soliton Solutions to The BS Equation]

\author{YU Song-Ju\ftnote{1}{fpc30017@bkc.ritsumei.ac.jp}, 
Kouichi TODA\ftnote{2}{sph20063@bkc.ritsumei.ac.jp},
Narimasa SASA\ftnote{3}{sasa@sugar.tokai.jaeri.go.jp} and 
Takeshi FUKUYAMA\ftnote{4}{fukuyama@bkc.ritsumei.ac.jp}}

\address{\dag \ddag Department of Physics, Ritsumeikan 
University, Kusatsu, Shiga 525 JAPAN}
\address{\S Center for Promotion of Computational Science and 
Engineering Japan Atomic Energy Research Institute, Tokai-mura, 
Ibaraki, 319-11, Japan}

\begin{abstract}
We study the integrable systems in higher dimensions which  can be 
written not by the Hirota's bilinear form but by the trilinear form.  
We explicitly discuss about the Bogoyavlenskii-Schiff(BS) equation 
in ($2 + 1$) dimensions.  Its analytical proof of multi soliton 
solution and a new feature are given.  Being guided by the strong 
symmetry, we also propose a new equation in ($3 + 1$) dimensions. 
\end{abstract}

\maketitle

\section{Introduction}

~~~~The Hirota's direct method (hereafter the direct method) 
provides us with a very powerful tool in the integrable 
systems\cite{rhirota}. Nakamura applied the direct method to the 
Ernst equation and obtained Tomimatsu-Sato solution(TS solution) 
in bilinear forms\cite{anakamura}.  However, his bilinear form does 
not take completely the same form as the conventional bilinear forms 
in the following senses.  It can not be expressed only by the 
Hirota's derivatives but involves ordinary derivatives.  Also the 
coefficients of the Hirota's derivatives are not constant but 
functions of independent variables.  Therefore, it was not trivial 
that the direct method works well in this system.  In the previous 
paper, we proved that the direct method does work in this 
system\cite{fky}.  However, our proof was complete in the restricted 
one dimensional case, Weyl solution, and was incomplete in full two 
dimensional case, TS solution.  Naive Pfaffian identity, which was 
valid for one dimensional case can not be applicable to double 
Wronskian in two dimensional case.  We consider the origin of this 
trouble lies in the peculiarities of the bilinear form mentioned above.  
By adopting the multilinear form\cite{grh,hgr}, we can rewrite the 
above bilinear forms so as to involve only multilinear operators.  
Thus we are forced to go beyond the bilinear form.  However, so far, 
any trilinear equations have not been shown to be integrable 
explicitly.  In this paper, we prove the integrability of the 
Bogoyavlenskii-Schiff(BS) equation\cite{b1,b2,jschiff}.  
Furthermore, being guided by the strong symmetry\cite{ff}, we search 
an integrable system in ($3 + 1$) dimensions.  

This paper is organized as follows.  In Sec.2, we construct the exact 
$N$ soliton solution of the BS equation in $N \times N$ Wronskian 
representation.  In Sec.3, a constructive proof of the $N$ soliton 
solution is given from the Miura transformation and the Hirota 
condition.  In Sec.4, we propose a new equation in ($3 + 1$) 
dimensions by the strong symmetry and give the travelling solitary 
wave solution to this system. Sec.5 is devoted to discussions.

\section{Exact $N$ Soliton Solution of the BS Equation in 
$N \times N$ Wronskian Representation}

~~~~We review the treatment to find the exact solutions of the KdV 
equation in the direct method for later use.  The KdV equation is 
written as

\begin{equation}
u_t + \Phi(u) u_x = 0, \label{kdv}
\end{equation}
where $\Phi(u) (\equiv \partial^2_x + 4 u + 2 u_x \partial^{-1}_x)$ 
is the strong symmetry\cite{ff}. Potential form of this equation is

\begin{equation}
\phi_{xt} + \phi_{4x} + 6 \phi_x \phi_{xx} = 0 \hspace{0.5cm} 
(u \equiv \phi_x) \label{pkdv}.
\end{equation}
By the dependent variable transformation

\begin{equation}
\phi \equiv 2 \frac{\tau_x}{\tau}, \label{dvtf}
\end{equation}
equation (\ref{pkdv}) is transformed into the bilinear form

\begin{equation}
\D_x (\D_t + \D_x^3) \tau \cdot \tau = 0, \label{bilinearkdv}
\end{equation}
where the Hirota's derivative $\D$ operating on $f \cdot g$ is 
defined by

\begin{equation}
\D^n_z f(z) \cdot g(z) \equiv (\partial_{z_1} - 
\partial_{z_2})^n f(z_1) g(z_2) \mid_{z_1 = z_2 = z}. \label{defdop}
\end{equation}
We have, in general, an exact solution $\tau_N$ which 
can be expressed as 

\begin{eqnarray}
& &\tau_N = 1 + \sum^N_{n = 1} \sum_{{}_NC_n} \eta_{i_1 \cdots i_n} 
\exp(\lambda_{i_1} + \cdots + \lambda_{i_n}), \label{nssofkdv} \\
& &\lambda_j = p_j x + \omega_j t + c_j, \hspace{0.5cm} \omega_j = - 
p^3_j, \label{etakdv} \\
& &\eta_{jk} = \frac{(p_j - p_k)^2}{(p_j + p_k)^2}, \label{ajk} \\
& &\eta_{i_1 \cdots i_n} = \eta_{i_1,i_2} \cdots \eta_{i_1,i_n} 
\cdots \eta_{i_{n - 1},i_n}. \label{aijklm}
\end{eqnarray}
where ${}_NC_n$ indicates summation over all possible combinations 
of $n$ elements taken from $N$, and symbols $c_j$ always denote 
arbitrary constants.  Equation (\ref{nssofkdv}) together with 
$u = 2 (\log \tau)_{xx}$ gives $N$ soliton solution of the KdV 
equation\cite{rhirota}. 

Then we proceed to the study of the BS equation which can be 
described not by the bilinear  form but by the trilinear form.  
The BS equation is given by 

\begin{equation}
u_t + \Phi(u) u_z = 0, \label{bs}
\end{equation}
Here $\Phi(u)$ has the same form as that in equation (\ref{kdv}) 
with argument $x$.  Using the potential $u \equiv \phi_x$, this 
equation reads

\begin{equation}
\phi_{xt} + \phi_{xxxz} + 4 \phi_x \phi_{xz} + 
2 \phi_{xx} \phi_z = 0 . 
\label{pbs}
\end{equation}
This equation has been constructed by Bogoyavlenskii and Schiff in 
the different ways.  Namely Bogoyavlenskii used the modified Lax 
formalism\cite{b1,b2}, whereas Schiff obtained the same equation 
by the reduction of the self dual Yang-Mills equation\cite{jschiff}.  
In reference \cite{grh,hgr,jschiff}, it was shown that equation 
(\ref{pbs}) is transformed into the trilinear form

\begin{equation}
\T_x (\T^3_x \T^{\ast}_z + 8 \T_x^2 \T^{\ast}_x \T_z + 9 \T_x \T_t) 
\tau \cdot \tau \cdot \tau = 0, \label{tribs}
\end{equation}
through the dependent variable transformation(\ref{dvtf}). The 
operators $\T$, $\T^{\ast}$ are defined by\cite{grh,hgr}

\begin{equation}
\fl \T_z^n f(z) \cdot g(z) \cdot h(z) \equiv (\partial_{z_1} + 
j \partial_{z_2} + j^2 
\partial_{z_3})^nf(z_1)g(z_2)h(z_3)|_{z_1=z_2=z_3=z}, \label{Tdef}
\end{equation}
where $j$ is the cubic root of unity, $j = \exp(2 \i \pi/3)$. 
$\T^{\ast}_z$ is the complex conjugate operator of $\T_z$ obtained 
by replacing $(\partial_{z_1} + j \partial_{z_2} + j^2 
\partial_{z_3})$ by $(\partial_{z_1} + j^2 \partial_{z_2} + j 
\partial_{z_3})$.
To find the $N$ soliton solutions, we repeat the same procedure as 
in the case of the KdV equation.  We find that $\tau_N$ is expressed 
as 

\begin{equation}
\tau_N = 1 + \sum^N_{n = 1} \sum_{{}_NC_n} \eta_{i_1 \cdots i_n} 
\exp(\lambda_{i_1} + \cdots + \lambda_{i_n}), \label{nssofbs}
\end{equation}
where

\begin{equation}
\lambda_j = p_j x + q_j z + r_j t + c_j, \hspace{0.5cm} 
r_j = - p^2_j q_j. \label{etabs}
\end{equation}
The proof is referred to Sec.3.  In the case of $N= 2$, the above 
2-soliton solution is same as that obtained by Schiff\cite{jschiff}.  

We rewrite $\tau_N$ in the form of $N \times N$ Wronskian, 

\begin{equation}
\tau_N = \det \pmatrix{
  f_1                    & \cdots  & f_N        \cr
  \vdots                 & \ddots  & \vdots     \cr
  \partial^{N - 1}_x f_1 & \cdots  & \partial^{N - 1}_x f_N   \cr
                                      }, \label{nsswronofbs}
\end{equation}
where

\begin{equation}
\fl f_j = \exp\biggl[\frac{1}{2} (p_j x + q_j z + r_j t + 
c_j)\biggr] + \exp\biggl[- \frac{1}{2} (p_j x + q_j z + r_j t + 
c_j)\biggr]. \\ 
\label{elements}
\end{equation}
The degree of variables in typical soliton equations are fixed.  
For example, the KdV equation (\ref{bilinearkdv}) demands that 

\begin{equation}
3 [\partial_x] = [\partial_t], \label{degkdveq}
\end{equation}
where  [$\partial_x$] is the degree of $\partial_x$.  So we may 
 set $[\partial_x] = 1$,

\begin{equation}
[\partial_x] = 1, \hspace{0.5cm} [\partial_t] = 3. \label{degkdveqs}
\end{equation}
We can use the Wronskian technique for the Wronskian solutions of 
the KdV equation\cite{hos}.  However, it is not the case in the BS 
equation.  Since equation (\ref{tribs}) only demands  

\begin{equation}
2 [\partial_x] + [\partial_z] = [\partial_t], \label{degbseq}
\end{equation}
equation (\ref{degbseq}) allows an indefinite factor, say $\alpha$, 
like 

\begin{equation}
[\partial_x] = 1, \hspace{0.5cm} [\partial_z] = \alpha,
\hspace{0.5cm} [\partial_t] = 2 + \alpha. \label{degbss}
\end{equation}
In this case, we cannot use the Wronskian technique by the 
presence of an indefinite factor $\alpha$.  This may enforce us 
to extended the Pfaffian identities.  We checked that 
(\ref{nsswronofbs}) are solutions to equation (\ref{tribs}) for an 
arbitrary $\alpha$ by the computer program Mathematica to $N = 8$.   

Figure \ref{1ssu} shows an example of the propagation of one 
soliton($u$).  The potential($\phi$) corresponding to Figure 
\ref{1ssu} with two floors is shown in Figure \ref{1ssp}.  In 
Figure \ref{2ss}, typical patterns of two solitons($p_1 \ne p_2$) 
and the potential with four floors are depicted.  In the soliton 
collision, however, appears a new feature.  For the special 
momentum combination ($p_1 = p_2 \ne 0$) two solitons shrink to V 
form(Figure \ref{2ssvsoli}): we may call this pattern V soliton.  

V soliton is a peculiar feature of the BS equation.  So let us 
discuss about it in more detail.  In the KP equation

\begin{equation}
\biggl(-4 u_t + \Phi(u) u_x\biggr)_x + 3 u_{yy} = 0, \label{kpeq}
\end{equation}
the resonance condition,

\begin{equation} 
\omega(\bf{k}_3)=\omega(\bf{k}_1) \pm \omega (\bf{k}_2),
\label{rescondition1}
\end{equation}
and 

\begin{equation} 
\bf{k}_3 = \bf{k}_1 \pm \bf{k}_2, 
\label{rescondition}
\end{equation}
with ${\bf k}_j \equiv (p_j,q_j)$ gives\cite{ky}, 

\begin{equation}
\fl (p_1 \pm p_2)^4 - 4 (p_1 \pm p_2) (\omega_1 \pm \omega_2) + 
3 (q_1 \pm q_2)^2 = \pm 3 p_1 p_2 \biggl((p_1 \pm p_2)^2 - 
(l_1 - l_2)^2\biggr) = 0, \\
\label{kprescondition}
\end{equation}
where $l_j \equiv q_j/p_j$.  Here $\tau_2 = 1 + e^{\lambda_1} + 
e^{\lambda_2} + \eta_{12}e^{\lambda_1 + \lambda_2}$ with 
$\lambda_j = p_j x + q_j y + \omega_j t + c_j$ ($p_j^4 - 
4 p_j \omega_j + 3 q_j^2 = 0$) and the phase shift $\eta_{12}$ is

\begin{equation}
\fl \eta_{12}= - \frac{(p_1 - p_2)^4 - 4 (p_1 - p_2) (\omega_1 - 
\omega_2) + 3 (q_1 - q_2)^2}{(p_1 + p_2)^4 - 4 (p_1 + p_2) 
(\omega_1 + \omega_2) + 3 (q_1 + q_2)^2}= \frac{(p_1 - p_2)^2 - 
(l_1 - l_2)^2}{(p_1 + p_2)^2 - (l_1 - l_2)^2}. \\ \label{kpaij}
\end{equation}
So the resonance condition corresponds to $\eta_{12}=0$ or $\infty$.  
In the BS equation the resonance condition (\ref{rescondition}) gives 

\begin{equation}
\frac{l_2}{l_1} = \mp \frac{p_2 \pm 2 p_1}{p_1 \pm 2 p_2}, 
\label{bsrescondition} 
\end{equation}
and $\eta_{12}$ is 

\begin{equation}
\eta_{12} = (\frac{p_1 - p_2}{p_1 + p_2})^2. \label{bsphaseshft}
\end{equation}
Thus the resonance condition corresponds to neither $\eta_{12} = 0$ 
nor $\eta_{12} = \infty$. To $\eta_{12} = 0$ corresponds V soliton.  
The soliton properties of V soliton are seen from the collision 
process of two V solitons. Two V solitons suffer phase shifting but 
conserve their solitary forms after collision. 

\section{Analytical Proof of $N$ Soliton Solutions to the BS 
Equation}

~~~~We give the analytical proof that equation(\ref{nssofbs}) is 
the solution to the BS equation (\ref{pbs}).  Firstly we introduce 
the modified Bogoyavlenskii-Schiff(mBS) equation which is deduced 
from the Miura transformation\cite{b2}.  This transformation 
connects the BS solution with the mBS solution.  The mBS equation 
is described by the coupled bilinear forms and tractable in the 
conventional Direct method.  Nextly we prove the integrability of 
the mBS equation.  This complete the proof of the BS solution.

Now we proceed to the concrete explanations. We perform the Miura 
transformation in the dependent variable of the BS 
equation(\ref{pbs}) . 

\begin{equation}
\phi_x = v^2 + \sigma v_x \hspace{0.5cm} (\sigma = \pm 1). 
\label{miuratf}
\end{equation}
Then we obtain the mBS equation,

\begin{equation}
v_t - 4 v^2 v_z - 2 v_x \partial^{- 1}_x ( v^2 )_z + v_{xxz} = 0. 
 \label{mbs} 
\end{equation}
Equation(\ref{mbs}) is reduced to the modified KdV equation in 
the case of $x = z$.  Introducing the new dependent variable 
$\psi$ by $v=\psi_x$ (\ref{mbs}), equation(\ref{mbs}) is reduced 
to the potential mBS equation

\begin{equation}
\psi_t - 2 \psi_x \partial^{- 1}_x ( \psi_x^2 )_z + \psi_{xxz} 
= 0. \label{pmbs}
\end{equation}
In order to eliminate the operator $\partial^{- 1}_x$ we describe 
this equation in terms of the coupled system,

\begin{eqnarray}
& &\rho_{xx} + \psi_x^2 = 0,  \label{pmbs1} \\
& &\psi_t + 2 \psi_x \rho_{xz} + \psi_z \rho_{xx} + 
\psi_x^2 \psi_z + \psi_{xxz} = 0. \label{pmbs2} 
\end{eqnarray}
Eliminating $\rho$, it is easily checked that equation(\ref{pmbs1}) 
and equation(\ref{pmbs2}) are equivalent to equation(\ref{pmbs}).
Here we perform the transformation of the dependent variables,

\begin{eqnarray}
& &\psi \equiv \log \biggl(\frac{F}{G}\biggr), \label{deptf1} \\
& &\rho \equiv \log (FG),  \label{deptf2}
\end{eqnarray}
then equations(\ref{pmbs1}), (\ref{pmbs2}) are reduced to the 
bilinear form,

\begin{eqnarray}
& &\D_x^2 F \cdot G = 0, \label{bilinearmBS1} \\
& &(\D_t + \D_x^2 \D_z) F \cdot G = 0. \label{bilinearmBS2}
\end{eqnarray}
$N$ soliton solutions of equations(\ref{bilinearmBS1}), 
(\ref{bilinearmBS2}), which we denote $F_N$, $G_N$ are speculated
from the conventional Hirota's Direct Method,

\begin{eqnarray}
& & F_N = 1 + \sum^N_{n = 1} \sum_{{}_NC_n} \eta_{i_1 \cdots i_n} 
\exp(\lambda_{i_1} + \cdots + \lambda_{i_n}), \label{nssofF} \\
& &G_N = 1 + \sum^N_{n = 1} \sum_{{}_NC_n} ( -1 )^n \eta_{i_1 \cdots i_n} 
\exp(\lambda_{i_1} + \cdots + \lambda_{i_n}), \label{nssofG} 
\end{eqnarray}
where $F_N$ is the same $N$ soliton solution of the BS 
equation(\ref{nssofbs}).  The proof is due to the Hirota 
condition\cite{rhirota2}.  We can rewrite the bilinear mBS 
equations(\ref{bilinearmBS1}), (\ref{bilinearmBS2}) as follows,

\begin{eqnarray}
& &\D_x^2 \tilde{f}_N \cdot \tilde{f}_N^{\ast} = 0, 
\label{newbilinearmBS1} \\
& &(\D_t + \D_x^2 \D_z) \tilde{f}_N \cdot \tilde{f}_N^{\ast} = 0. 
\label{newbilinearmBS2}
\end{eqnarray}
Here

\begin{eqnarray}
& &\tilde{f}_N = \sum^N_{\underline{\mu} = 0, 1} \exp\biggl(\sum_{j=1}^N 
\mu_j (\lambda_j + i \frac{\pi}{2}) + \sum_{1 \le j < k}^N \mu_j \mu_k 
A_{jk}\biggr), \label{nssoftildef} \\
& &\tilde{f}_N^{\ast} = \sum^N_{\underline{\nu} = 0, 1} 
\exp\biggl(\sum_{j=1}^N \nu_j (\lambda_j - i \frac{\pi}{2}) + 
\sum_{1 \le j < k}^N \nu_j \nu_k A_{jk}\biggr), \label{nssoftildefast} \\
& &\exp(A_{jk}) \equiv \eta_{jk} = \frac{(p_j - p_k)^2}{(p_j + p_k)^2}. 
\label{expa}
\end{eqnarray}
$\sum^N_{\underline{\mu}}$, $\sum^N_{\underline{\nu}}$ denote the 
summation of $\mu_j = 0,1$, $\nu_j = 0,1$($j = 1, 2, \cdots, N$).  
Substitution of the expression for $\tilde{f}_N$ and 
$\tilde{f}_N^{\ast}$ into equations(\ref{newbilinearmBS1}), 
(\ref{newbilinearmBS2}) reveals that the coefficients of 
$\exp(\sum_j^n \lambda_j + \sum_{j = n + 1}^m 2 \lambda_j)$ are all 
vanished for the respective $n$ and $m$, 

\begin{eqnarray}
\fl & & \sum^N_{\underline{\mu} = 0, 1} \sum^N_{\underline{\nu} = 0, 1} 
\Biggl(\biggl(\sum_{j = 1}^N (\mu_j - \nu_j) p_j\biggr)^2\Biggr) 
\exp\biggl(\sum_{j = 1}^N \frac{i \pi}{2} (\mu_j - \nu_j) 
+ \sum_{1 \le j < k}^N (\mu_j \mu_k + \nu_j \nu_k) A_{jk}\biggr) 
\nonumber \\
\fl &\times& \mbox{cond}(\underline{\mu},\underline{\nu})_{nm} = 0, 
\label{newnewbilinmBS1} \\
\fl & & \sum^N_{\underline{\mu} = 0, 1} \sum^N_{\underline{\nu} = 0, 1} 
\Biggl(\biggl(\sum_{j = 1}^N (\mu_j - \nu_j) p_j\biggr)^2 
\biggl(\sum_{j = 1}^N (\mu_j - \nu_j) q_j\biggr) 
- \biggl(\sum_{j = 1}^N (\mu_j - \nu_j) p_j^2 q_j\biggr)\Biggr) 
\nonumber \\ 
\fl &\times& \exp\biggl(\sum_{j = 1}^N \frac{i \pi}{2} (\mu_j - \nu_j) 
+ \sum_{1 \le j < k}^N (\mu_j \mu_k + \nu_j \nu_k) A_{jk}\biggr) 
\mbox{cond}(\underline{\mu},\underline{\nu})_{nm} = 0,  
\label{newnewbilinmBS2}
\end{eqnarray}
where

\begin{equation}
\fl \mbox{cond}(\underline{\mu},\underline{\nu})_{nm} = \left\{
   \begin{array}{rl}
	1, & \quad \mbox{for $j = 1$, $\cdots$, $n$ : 
$\mu_j + \nu_j = 1$, $0 \le n \le N$} \\
	1, & \quad \mbox{for $j = n + 1$, $\cdots$, $m$ : 
$\mu_j = \nu_j = 1$, $n \le m \le N$} \\
	1, & \quad \mbox{for $j = m + 1$, $\cdots$, $N$ : 
$\mu_j = \nu_j = 0$} \\
	0, & \quad \mbox{otherwise}
   \end{array}\right.
\end{equation}
Here we have used the notations by J. Ablowitz and H. Segur
\cite{as}.  As is easily seen, the first case of 
$\mbox{cond}(\underline{\mu},\underline{\nu})_{nm}$ gives the non 
trivial contribution. Equations(\ref{newnewbilinmBS1}) and 
(\ref{newnewbilinmBS2}) are reduced to the following equations
(\ref{hirotacondition1}) and (\ref{hirotacondition2}), 
respectively for a given $n$.

\begin{eqnarray}
\fl & &\sum^n_{\underline{\sigma} = \pm 1} 
\Biggl(\biggl(\sum_{j = 1}^n \sigma_j p_j\biggr)^2\Biggr) 
\exp\biggl(\frac{i \pi}{2} \sum_{j = 1}^n \sigma_j\biggr) 
\prod_{j < k}^n (\sigma_j p_j - \sigma_k p_k)^2 = 0, 
\label{hirotacondition1} \\
\fl & &\sum^n_{\underline{\sigma} = \pm 1} 
\Biggl(\biggl(\sum_{j = 1}^n \sigma_j p_j\biggr)^2 
\biggl(\sum_{j = 1}^n \sigma_j q_j\biggr) - 
\biggl(\sum_{j = 1}^n \sigma_j p_j^2 q_j\biggr)\Biggr) 
\exp\biggl(\frac{i \pi}{2} \sum_{j = 1}^n \sigma_j\biggr) \nonumber \\
\fl &\times& \prod_{j < k}^n (\sigma_j p_j - \sigma_k p_k)^2 = 0, 
\label{hirotacondition2}
\end{eqnarray}
where

\begin{equation}
\sigma_j \equiv \mu_j - \nu_j. \label{sigma}
\end{equation}
Equation(\ref{hirotacondition1}) is easily verified for $n = 1, 2$.  
Let as denote the left-hand side of equation(\ref{hirotacondition1}) 
as $\tilde{\triangle}(n)$ .  Then $\tilde{\triangle}(n)$ has the 
following properties:(i) $\tilde{\triangle}(n)$ is a symmetric 
homogeneous polynomial of $p_j$, (ii) if $p_1 = 0$ then 
$\tilde{\triangle}(n) = 0$, (iii) if $p_1 = p_2$ then 

\begin{equation}
\tilde{\triangle}(n) = 4 p_1^2 \prod_{k = 3}^n 
(p_1^2 - p_k^2)^2 \tilde{\triangle}(n-2).
\end{equation}
Now we assume that equation(\ref{hirotacondition1}) holds for 
$n - 2$.  Then, using the properties (i), (ii) and (iii), we find 
that $\tilde{\triangle}(n)$ can be factored by a symmetric 
homogeneous polynomial

\begin{equation}
\prod_{j = 1}^n p_j \prod_{1 \le j < k}^n (p_1^2 - p_k^2)^2, \label{skata}
\end{equation}
of degree $n^2$. On the other hand, equation(\ref{hirotacondition1}) 
shows the degree of $\tilde{\triangle}(n)$ to be $n^2 - n + 2$.  
Hence, $\tilde{\triangle}(n)$ must be zero for $n$.

Next we discuss equation(\ref{hirotacondition2}).  we can rewrite 
equation(\ref{hirotacondition2}) as  

\begin{equation}
\sum_{j = 1}^n q_j \tilde{\triangle}_j(n) = 0, \label{2tilda}
\end{equation}
where equation(\ref{2tilda}) is a symmetric homogeneous polynomial of 
($p_j$,$q_j$).

\begin{eqnarray}
\fl  \tilde{\triangle}_1(n) &=& \sum^n_{\underline{\sigma} = \pm 1} 
\Biggl(\biggl(\sum_{j = 1}^n \sigma_j p_j\biggr)^2 \sigma_1 - 
\biggl(\sum_{j = 1}^n \sigma_1 p_1^2\biggr)\Biggr) 
\exp\biggl(\frac{i \pi}{2} \sum_{j = 1}^n \sigma_j\biggr) 
\prod_{j < k}^n (\sigma_j p_j - \sigma_k p_k)^2 \nonumber \\
\fl &=& 2 i \sum_{\sigma_2 = \pm 1, \cdots, \sigma_n = \pm 1} 
\Biggl(- 4 p_1^2 i^{n - 1} \biggl(\prod_{j = 2}^n p_j\biggr) 
\biggl(\sum_{j = 2}^n \sigma_j p_j\biggr) \prod_{2 \le j < k}^n 
(\sigma_j p_j - \sigma_k p_k)^2 \nonumber \\
\fl &+& \biggl(\prod_{j = 2}^n (p_1^2 + p_j^2)\biggr) 
\biggl(\sum_{j = 2}^n \sigma_j p_j\biggr)^2 \exp\biggl(\frac{i \pi}{2} 
\sum_{j = 2}^n \sigma_j\biggr) \prod_{2 \le j < k}^n 
(\sigma_j p_j - \sigma_k p_k)^2\Biggr), \label{1tilda}
\end{eqnarray}
etc.  The first term of the right-hand side of 
equation(\ref{1tilda}) must be zero because this term contains 
only the odd powers of each $\sigma_j$($j = 2, \cdots, n$), the 
second term equal to zero from equation(\ref{hirotacondition1}).  
Hence, equation(\ref{hirotacondition2}) holds.

Therefore equation(\ref{nssofbs}) is the soliton solution of the 
BS equation from the Miura transformation(\ref{miuratf}).  This 
completes the proof.

\section{A New Equation in ($3 + 1$) Dimensions and Its Travelling 
Solitary Wave Solutions}

~~~~We have studied how the KdV equation in ($1 + 1$) dimensions is 
extended to the KP equation and the BS equation in ($2 + 1$) 
dimensions.  Namely, we have two different ways to the integrable 
systems in one higher dimensions.  So further analogy leads us to 
the new systems in two higher dimensions, ($3 + 1$) dimensions, 

\begin{equation}
\biggl(-4 u_t + \Phi(u) u_z\biggr)_x + 3 u_{yy} = 0. \label{neweq}
\end{equation}
These extension schemes are schematically written in the following 
form: 

\vspace{0.5cm}
\begin{center}
\fbox{\parbox{7.5cm}{{\bf
KdV equation (\ref{kdv}) $\Longrightarrow$ BS equation (\ref{bs})

\hspace{1cm}$\Downarrow$ \hspace{3.5cm}$\Downarrow$
 
KP equation (\ref{kpeq}) $\Longrightarrow$ Equation (\ref{neweq})
}}}
\end{center}
\vspace{0.5cm}
Equation (\ref{neweq}) was expected to be integrable. However, 
the potential form of equation (\ref{neweq}), 

\begin{equation}
-4 \phi_{xt} + \phi_{xxxz} + 4 \phi_x \phi_{xz} + 2 \phi_{xx} \phi_z 
+ 3 \phi_{yy} = 0, \hspace{0.5cm} (u \equiv \phi_x) \label{neweqpot}
\end{equation}
has a movable logarithmic branch point in the 
sense of WTC method\cite{wtc}.  Furthermore, we can not construct 
$N (\ge 2)$ soliton solution of trilinear form of equation 
(\ref{neweq}) 

\begin{equation}
(\T^4_x \T^{\ast}_z + 8 \T^3_x \T^{\ast}_x \T_z - 36 \T^2_x \T_t + 
27 \T_x \T^2_y) \tau \cdot \tau \cdot \tau = 0, \label{neweqtri}
\end{equation}
by the direct method.  We require the existance of $2$ soliton 
solution. If $2$ soliton solution,

\begin{eqnarray}
& & \tau_2 = 1 + \exp(\lambda_1) + \exp(\lambda_2) \eta_{12} 
\exp(\lambda_1 + \lambda_2), \label{2ss3+1eq} \\
& & \lambda_j \equiv p_j x + q_j y + r_j z + s_j t + c_j. 
\label{lambda3+1eq}
\end{eqnarray}
then

\begin{eqnarray}
& & \eta_{12} = \frac{\alpha p_1^2 p_2^2 (p_1 - p_2)^2 
- (q_1 p_2 - q_2 p_1)^2}{\alpha p_1^2 p_2^2 (p_1 + p_2)^2
- (q_1 p_2 - q_2 p_1)^2}, \label{eta3+1eq} \\
& & r_1 = \alpha p_1, \hspace{0.5cm} r_2 = \alpha p_2, \label{condition3+1eq}
\end{eqnarray}
where $\alpha$ is arbitrary constant, thus equation(\ref{neweqtri}) 
is reduced to ($2 + 1$) dimensional equation.  These suggest that 
equation (\ref{neweq}) is not integrable.  However, 
equation(\ref{neweq}) has explicit travelling solitary wave solution 
by $\tanh$-function method(TFM)\cite{pd}.  The ansatz is expressible 
as a polynomial in terms of a $\tanh$ function, so that it has the 
form

\begin{equation}
u(x,y,z,t) = U(\eta) = \sum_{i = 0}^M a_i T^i, \hspace{0.5cm} T \equiv 
\tanh (k \eta), \label{ansatz}
\end{equation}
where $\eta = x + l y + m z - c t + {\rm constant}$. Substitution 
of equation (\ref{ansatz}) into equation (\ref{neweq}) yields an 
ordinary differential equation for $U(\eta)$

\begin{equation}
(4 c + 3 l^2) U + 3 m U^2 + m \frac{d^2 U}{d \eta^2} = b, \label{odeneweq}
\end{equation}
where $b$ is an integrable constant. 

We balance the highest power of $T$ in the second term in equation 
(\ref{odeneweq}) with the highest power of $T$ in the final term 
in equation (\ref{odeneweq}) to obtain $2 M = M + 2$, so that 
$M = 2$.  In order to solve equation (\ref{odeneweq}) we use the 
automated tanh-function method(ATFM)\cite{pd}, where one inputs the 
commands in {\it Mathematica}, and obtain the outputs in the 
following ways:

\begin{verbatim}
In[1]:= << atfm`
In[2]:= neweq = (4 c + 3 l^2) U[T] + 3 m U[T]^2 + m der[U[T],T,2] - b;
In[3]:= ATFM[neweq, U, T, 2, c, l, m, b]
                  2
{a[0] + T a[1] + T  a[2], k, c, l, 0, 0}
    2          2                         
 4 k    2 c   l        2  2              
{---- - --- - --- - 2 k  T , k, c, l, m, 
  3     3 m   2 m                        

       2         2      4       4  2
  -16 c  - 24 c l  - 9 l  + 16 k  m
  ----------------------------------},
                 12 m
\end{verbatim}
which shows the solution

\begin{equation}
\fl u(x,y,z,t) = \frac{4 k^2}{3} - \frac{2 c}{3 m} - \frac{l^2}{2 m} - 
2 k^2 \tanh^2 \biggl(k (x + l y + m z - c t + d)\biggr). \\
\label{solofneweqtype1}
\end{equation}
Here $c$, $d$, $k$, $l$ and  $m$ are arbitrary constants, and $b$ 
becomes 

\begin{equation}
b = \frac{- 16 c^2 -24 c l^2 - 9 l^4 + 16 k^4 m^2}{12 m}. \label{b}
\end{equation}
Note that $b$ should vanish for soliton solution in which $u \to 0$ 
as $|\eta| \to \infty$. In this case equation (\ref{b}) is reduced 
to

\begin{equation}
3 l^2 + 4 c = \pm 4 k^2 m. \label{solicon}
\end{equation}
Substitution of the choice $3 l^2 + 4 c = - 4 k^2 m$ into the 
solution(\ref{solofneweqtype1}) gives the familiar ${\rm sech}^2$ 
solution, 

\begin{equation}
u(x,y,z,t) = 2 k^2 {\rm sech}^2 \biggl(k (x + l y + m z - c t + 
d)\biggr). \label{solofneweqtype2}
\end{equation}
\section{Discussions}

~~~~In this paper, we have obtained the exact $N$ soliton solution 
of the BS equation and the travelling solitary wave solution of 
equation (\ref{neweq}).  These two solutions seem to have 
essentially the same structure as that of the KdV equation.  Indeed 
their spatial dependences are described by a new single variable 
like $p_j x + q_j y = p_j' x'$ in equation (\ref{elements}) and 
$x + l y + m z = x'$ in equation (\ref{solofneweqtype2}).  However 
if we consider multi soliton solution and multi soliton collision, 
the extra dimensions plays essential roles and complex the analytical 
proof of $N$ soliton solutions.  V soliton is one of such examples. 
If we remark V soliton collision on some spatial axis we see that 
two solitons in ($1 + 1$) dimension come together and disappear or 
that two solitons come to birth from nothing.  This does not occur 
in the KdV equation. It is worth noting that this latter process 
occurs in the Broer-Kaup equation which is the ($1+1$) dimensional 
integrable system written in the trilinear form \cite{ms,skmh}.  

Our treatment of extension of integrable system to higher 
dimensions indicates some analogy to that of the $d$ dimensional  
cylindrical KdV equation.  The latter system is described by

\begin{equation}
u_t + 6 u u_x + u_{xxx} + \frac{(d-1)}{2 t} u=0, \label{hkdv}
\end{equation}
where $d=1, 2$ and $3$ correspond to the KdV, the cylindrical 
KdV and the spherical KdV equations, respectively. The last term 
is the curvature term.  The Painlev\'e test indicates that $d=1$ 
and $2$ cases are integrable but that $d=3$ case has the movable 
branch point\cite{anaka,br,fy1}. This is the same situation as 
the KdV ($d=1$), the BS equation ($d=2$) and new equations 
($d=3$).  However, at last at this stage, it is not clear 
whether these resemblances have any deep implication or not.  
One of our future concerns is to construct an integrable system 
in ($3 + 1$) dimensions which is reduced to the BS equation and 
to the KP equation in some particular cases.

\ack

We would like to thank K. Kamimura, J. Schiff and K. Takasaki for 
helpful discussions.  We also acknowledge to E. J. Parkes and 
B. R. Duffy to sending $\mbox{ATFM}$ package.


\newpage

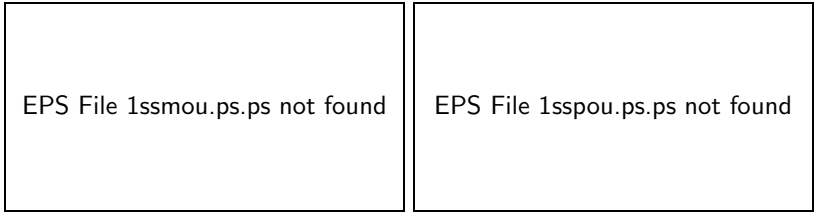
\begin{figure}
\begin{center}
  \epsfile{file=1ssmou.ps,scale=0.35}
  \epsfile{file=1sspou.ps,scale=0.35}
\end{center}
\caption{}
\label{1ssu}
\end{figure}

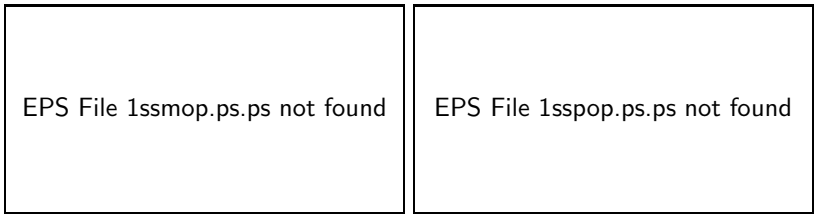
\begin{figure}
\begin{center}
  \epsfile{file=1ssmop.ps,scale=0.35}
  \epsfile{file=1sspop.ps,scale=0.35}
\end{center}
\caption{}
\label{1ssp}
\end{figure}

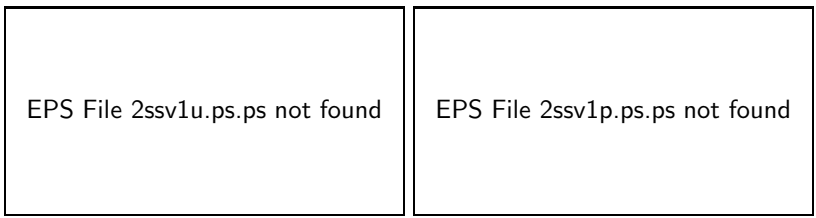
\begin{figure}
\begin{center}
  \epsfile{file=2ssv1u.ps,scale=0.35}
  \epsfile{file=2ssv1p.ps,scale=0.35}
\end{center}
\caption{\hspace{0.2cm} (a) \hspace{4cm} (b)}
\label{2ss}
\end{figure}

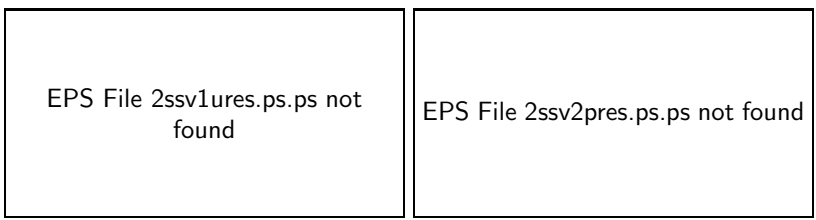
\begin{figure}
\begin{center}
  \epsfile{file=2ssv1ures.ps,scale=0.35}
  \epsfile{file=2ssv2pres.ps,scale=0.35}
\end{center}
\caption{\hspace{0.2cm} (a) \hspace{4cm} (b)}
\label{2ssvsoli}
\end{figure}

{\bf Figure Captions}:

\vspace{0.5cm}

\begin{enumerate}
\item[Figure \ref{1ssu}]: Time evolution of the one soliton solution 
$u$ with $p_1 = 2$, $q_1 = -3$.

\vspace{0.5cm}

\item[Figure \ref{1ssp}]: Time evolution of $\phi$ with $p_1 = 2$, 
$q_1 = -3$.

\vspace{0.5cm}

\item[Figure \ref{2ss}]: \begin{enumerate}
                         \item[(a)] An example of the two soliton 
                                    solution with $p_1 = 0.3$, 
                                    $p_2 = - 0.2$, $q_1 = - 0.15$,
                                    $q_2 = - 0.1$. 
                         \item[(b)] Potential diagram corresponding 
                                    to (a)
                         \end{enumerate}

\vspace{0.5cm}

\item[Figure \ref{2ssvsoli}]: \begin{enumerate}
                              \item[(a)] An example of the two soliton 
                                         solution with $p_1 = p_2 = 0.3$, 
                                         $q_1 = - 0.15$, $q_2 = 0.1$.
                              \item[(b)] Potential diagram corresponding 
                                         to (a)
                              \end{enumerate}
\end{enumerate}

\newpage

\end{document}